\documentstyle[12pt,aaspp4]{article}
\tightenlines

\newcommand{\kms}{\mbox{km s$^{-1}$}}
\newcommand{\etal}{{\it et al}.\ }

\newcommand{\lta}{\stackrel{<}{\sim}}

\newcommand{\gta}{\stackrel{>}{\sim}}
\begin{document}
\title{X-Ray Sources in Regions of Star Formation. \\ VI.
The R CrA Association as Viewed by {\it EINSTEIN}.}

\author{Frederick M. Walter\altaffilmark{1,2}}
\affil{Department of Physics and Astronomy, State University of New York, Stony
Brook NY 11794-3800\\I:fwalter@astro.sunysb.edu}

\author{Frederick J. Vrba\altaffilmark{2}}
\affil{U.S. Naval Observatory, Flagstaff Station, Flagstaff AZ 86002\\
I:fjv@nofs.navy.mil}

\author{Scott J. Wolk\altaffilmark{2}}
\affil{Earth and Space Sciences Department, State University of New York, Stony
Brook NY 11794-2100 and
Harvard-Smithsonian Center for Astrophysics, Cambridge MA 02138\\
I:swolk@head-cfa.harvard.edu}

\author{Robert D. Mathieu}
\affil{Department of Astronomy, University of Wisconsin, Madison WI 53706
and Harvard-Smithsonian Center for Astrophysics, Cambridge MA 02138\\
I:mathieu@astro.wisc.edu}

\author{Ralph Neuh\"auser}
\affil{Max-Planck-Institut f\"ur Extraterrestrische Physik,
       85740 Garching, Germany\\I:rne@hpth03.mpe-garching.mpg.de}

\altaffiltext{1}{Guest Observer, EINSTEIN Observatory.}
\altaffiltext{2}{Visiting Astronomer, Cerro Tololo Inter-American and Kitt
     Peak National
     Observatories, National Optical Astronomy Observatories, which are
      operated by the  Association of Universities for Research in Astronomy,
     Inc., under contract with the National Science Foundation.}

\begin{abstract}

We report on optical identifications of X-ray sources in the vicinity
of the R~CrA association. We identify 11 low-mass pre-main-sequence stars as
counterparts of 9 X-ray sources.
We also find X-ray emission from coordinates consistent with the position
of 3 bright late-B stars, although the emission may come from lower mass
companions. 
The X-ray-selected stars lie along a narrow locus in the
luminosity-temperature diagram at an age of about
7~Myr, which is considerably older than the estimated ages of the 
higher mass stars or of the IR-excess stars.
We determine the physical characteristics of these stars,
including masses, ages, and Lithium abundances.
We estimate that the complete membership of the R~CrA association amounts to
about 90 stars, mostly older naked T~Tauri stars, and that 
the population is consistent with the standard IMF.

\end{abstract}


\section{Introduction}
The R~CrA association, at a distance of about 130~pc \markcite{mr81}(Marraco
\& Rydgren 1981), is one of the nearest star-forming associations to the Sun.
The association is projected against a prominent dark cloud 18$^o$ below the
galactic plane (see figure~1 of \markcite{k73}Knacke \etal 1973). The total
mass of the cloud is estimated to be between 3000 and 10,000~M$_\odot$
\markcite{D87}(Dame \etal 1987; \markcite{l79}Loren 1979). \markcite{bw94}
Wilking \etal (1994) comment that the morphology of the cloud bears a
resemblance to the $\rho$~Oph cloud. However, the number of known and suspected
members of the stellar association is only 10\% of that known or suspected in
$\rho$~Oph. 

\markcite{k73}Knacke \etal (1973) discovered a group of 11 young stars near
R~CrA. Members of the association include the intermediate-mass stars R, T, and
TY~CrA and the classical T Tauri stars S, VV, and DG~CrA. Nine of the 11 stars
had near-IR continuum excesses. \markcite{k73}Knacke \etal\ also identified 10
H$\alpha$-emitting stars spread over the region, at projected distances up to a
half degree from the dark cloud. \markcite{gp75}Glass \& Penston (1975)
undertook near-IR photometry of 44 stars in this region, and noted that the
only stars with near-IR continuum excesses were the known variable stars. They
failed to detect near-IR excesses for 3 of the H$\alpha$-emitting stars
identified by \markcite{k73}Knacke \etal, and commented that H$\alpha$
emission and 2$\mu$m excesses do not necessarily correlate. \markcite{mr81}
Marraco \& Rydgren (1981) reexamined the association. They found H$\alpha$
emission in only 4 of the 10 Knacke \etal sources, but discovered another 6
H$\alpha$-emitting objects. \markcite{bw92} Wilking \etal (1992) surveyed this
region using the IRAS data, finding 16 IRAS sources associated with young
stellar objects (YSOs). They identified 24 YSOs from their near- and far-IR
spectral energy distributions. \markcite{bw94} Wilking \etal (1994) presented
results of near-IR imaging of the cloud core, and concluded that there are
approximately 30 members of this association. 

\markcite{k73} Knacke \etal (1973) estimated an association age of less than
10$^6$~years (1~Myr), assuming that the late-B stars of the association have
just arrived on the main-sequence. \markcite{bw92} Wilking \etal (1992)
estimate an age of between 1.5~Myr (the contraction age of the zero-age
main sequence B8 star
TY~CrA) and 6~Myr (the contraction time for the pre-main sequence A5 star
R~CrA). 

While infrared observations can be useful in identifying low-mass 
pre-main-sequence (PMS) stars,
experience over the past decade has shown that the
population sampled in the infrared is not necessarily representative of the 
true population of young associations. X-ray imaging observations (e.g.,
\markcite{fk81}Feigelson \& Kriss 1981; \markcite{m83}Montmerle \etal 1983; 
\markcite{f93}Feigelson \etal 1993;
\markcite{mu83}Mundt \etal 1983; \markcite{ss94}Strom \& Strom 1994;
\markcite{wa88}\markcite{wa94}Walter \etal 1988, 1994)
of young associations
reveal a population of X-ray-luminous young stars which generally
lack infrared continuum excesses. 
\markcite{wa86}Walter (1986) found 3 X-ray-luminous PMS stars near the
CrA cloud. 
Here we report on the complete EINSTEIN
Observatory Imaging Proportional Counter (IPC) observations of the CrA dark
cloud. In addition to the 3 sources previously identified, there are another
8 X-ray sources. In this paper we discuss the identifications of these X-ray
sources.

\section{Observations}
\subsection{X-Ray Observations}
The X-ray data consist of the two overlapping IPC observations listed in
Table~\ref{tbl-1} and shown in Figure~\ref{fig1}.
The IPC ribs, which support the detector window,
obscure $\sim$30\% of the nominal 1~square-degree field of the IPC.
There is significant vignetting in the IPC;
effective exposure times at the edge of the field (30~arcmin from the
center) are about 50\% of those on-axis.

The IPC is sensitive to photons with energies between about 0.1 and 4.5 keV.
Active stellar coronae have characteristic temperatures of 10$^7$K, or about
1~keV, and are well matched to the IPC response.
The energy resolution of the IPC is
about 100\%, and only gross spectral information is available for weak sources.
The 90\% confidence uncertainty on the X-ray source positions is
typically $\pm$30-40~arcseconds.
Details of the IPC are given in \markcite{g79}Giacconi \etal (1979).

There are three X-ray sources in IPC sequence 
3501, centered on $\epsilon$~CrA (HD175813). Two of these were found in the
standard rev~1B processing of sequence. The MDETECT (map detect)
source detection algorithm had not run as a part of the standard processing.
We requested that MDETECT be run off-line, and it yielded the third
X-ray source. 
In IPC sequence 4512, centered on the CrA dark
cloud, the standard processing yielded 10 X-ray sources.

\subsection{Optical and near-IR observations}

All of the X-ray sources had potential optical counterparts on the POSS
sky survey. The W~UMa star $\epsilon$~CrA
\markcite{cd84}(Cruddace \& Dupree 1984) was detected in both IPC
observations.  The other X-ray sources, their J2000 coordinates, and aliases,
are presented in Table~\ref{tbl-2}.

We obtained low-dispersion spectra of these stars at the
Kitt Peak National Observatory in June 1983, April 1986, and May 1987,
using the Intensified Image Dissector Scanner (IIDS) on the 2.1-m telescope.
All observations were made
through the 7-arcsecond aperture. Observations in the red covered
the range 6150-7200\AA\ at about 3\AA\ resolution, and were obtained with
grating~36 and the GG495 order-sorting filter. Blue observations, taken using
grating~36 and the CuSO$_4$ order-sorting filter, covered the
range 3750-4200\AA\ with about 1.5\AA\ resolution. The spectra were fully
reduced and wavelength- and flux-calibrated
using the mountain data reductions and the IPPS software package. Flux
calibration relied on the IRS standard stars 
\markcite{bh82}(Barnes \& Hayes 1982).
Subsequent data analysis was performed using software written in IDL.

Low-dispersion spectra were also taken on 18 and 23 July 1995, and on 21
July 1996,
at the European Southern Observatory (ESO) using the 1.52m telescope
equipped with a Boller and Chivens spectrograph. We
used grating \#5 and CCD \#24 (in 1995) or CCD \#39 (in 1996),
achieving a mean
resolution of about 2.5\AA\ (FWHM) in the 4600-7000\AA\ spectral range.
This instrumental set-up allows us to resolve the Lithium~I
absorption line at 6707\AA\ from the Ca~I absorption line at 6717\AA\ as
well as to carry out an accurate spectral type classification.
For the wavelength calibration, a He-Ar exposure
was taken after each science exposure. The standard stars LTT 4816,
LTT 7987, and EG 274 were observed each night with the same instrumental
set-up to perform the flux calibration.
We used the MIDAS software (versions 95MAY and 95NOV) to reduce
these spectra. Bias and dark subtraction was first performed on each frame.
The 2-D science frames were then divided by a mean flat-field and then
calibrated in wavelength. The sky subtraction, the extraction of one
dimensional spectra and the flux calibration using a mean response function,
were finally performed.

High-dispersion spectra were obtained using the Cerro Tololo Inter-American
Observatory (CTIO) 4-m echelle on 4-6 April 1987 and 23-27 May 1989. Using the
red Air Schmidt camera with the 31.6 l/mm echelle, cross-disperser \#3, the
200$\mu$m slit, and the GG495 order sorting filter, we obtained 17~\kms\
spectral resolution from 5600 to 7000\AA . We used decker 1 in April 1987 and
decker 2 in 1989. The CCD detector was EPI~\#9 in 1987 and EPI~\#12 in 1989. We
subtracted the overscan region and the CCD bias and trimmed the frames using
IRAF. Subsequent reductions were undertaken using an echelle data reduction
package written in IDL \markcite{wa92}(Walter 1992). The data were
flux-calibrated using Kopff~27 and LTT~3864 as flux-standard stars. Most
observations consisted of 3 sequential integrations; cosmic rays were filtered
out by applying a median filter to the multiple observations of each target.
The 1987 observations were compromised by unusually large flexure within the
spectrograph. In some cases flexure shifted the spectra off the region
illuminated by the projector flat (taken using decker 2), resulting in lower
S/N than indicated by counting statistics alone. Radial velocities obtained
during this run are unreliable, and have not been used. 

We obtained multiple high-dispersion single-order spectra of each star with the
echelle spectrographs of the Whipple Observatory 1.5-m reflector and the MMT
between 1986 and 1993 for the purpose of measuring radial velocities. All the
spectra were centered at $\lambda$5187\AA\ and provided about 50\AA\ coverage.
Radial velocities were obtained via cross-correlation, following the procedures
discussed by \markcite{n94}Nordstrom \etal (1994; see also
\markcite{l92}Latham 1992).
Cross-correlations were made against a grid of synthetic
templates with a separation of 250K in temperature, 0.5 in log g, and 10~\kms\
in V~{\it sin~i}. For each star the radial velocity was adopted for that
template providing the highest value for the average of the correlation peak.
The typical precision of a single radial-velocity measurement is 1~\kms , based
on the dispersions of repeated measurements. Rotational velocities were
determined through an interpolation based on correlation peak strength, with a
measurement lower limit of $\approx$ 10~\kms. We estimate an uncertainty of
5~\kms. 

UBVRI photometry was obtained in May 1986, June 1987, and May 1988 at CTIO. The
data were obtained with the 1.0-m telescope using a Hamamatsu GaAs
photomultiplier and apertures ranging from 15-20~arcseconds. All UBV
observations are on the \markcite{j63}Johnson (1963) system, while all R and I
observations are on the \markcite{c80}Cousins (1980) system. We observed
secondary standards from \markcite{al83}Landolt (1983). Additionally, a few
UBV obsevations were obtained for us by Dr.\ Robert Millis during May 1986 with
the CTIO/Lowell 0.4-m telescope using an EMI6256 photomultiplier. Multiple,
independent observations were obtained of each star during these observing runs
with no obvious systematic differences in results between the various systems
employed. 

Further UBVRI photometry was obtained from CCD images acquired at CTIO on 1993
May 5 and 1994 June 3 using the 0.9 meter telescope.  The detector used was the
TEK 2K-1 CCD with the standard Bessell filter set.  CCD debiasing and
flat-fielding were performed using the IRAF CCDRED package.  Instrumental
magnitudes were determined using the IRAF APPHOT package.  In the case of the
close pair CrAPMS~6, photometry was performed on the individual stars after
first subtracting off the other star using routines in the IRAF DAOPHOT
package. The photometry was placed on the standard Johnson-Kron-Cousins system
using calibration stars from \markcite{al92}Landolt (1992).  Mean residual
errors are less than 3\% except in the case of the CrAPMS~6 pair, which
individually have mean residual errors of about 5\%. 

We also obtained near-infrared JHKL photometry during June 1986 using the CTIO
4-m telescope and during May 1989 using the CTIO 1.5-m telescope. Both
observing runs employed a monolithic InSb detector and used standards from
\markcite{e82}Elias \etal (1982) to place the observations onto the standard
system. At the 4-m telescope a 6-arcsecond aperture was used, while at the
1.5-m telescope a 14-arcsecond aperture was used. One to five independent
observations were obtained of each program star. 

We obtained additional JHK observations on 19-21 February 1997, using the CIRIM
imaging detector on the CTIO 1.5-m telescope. Each image consisted of a raster
of 6 exposures, with the telescope offset by 15~arcsec between exposures. We
observed standard stars from \markcite{e82}Elias \etal (1982). The
non-linearity correction was applied using the IRAF task IRLINCOR. All
subsequent data reductions were undertaken using IDL software written for the
task. The images were flattened using dome-flats. The local sky was determined
by median-filtering the six images, and was subtracted off. The frames were
then registered and coadded. We used the standard star observations to
determine the extinction correction for each night and place the photometry on
the standard CIT system. Uncertainties in the photometric solution amount to
$\pm$0.02~mag for single observations. Target fluxes were measured in an
8~arcsec radius aperture. For close binaries we measured the relative fluxes in
smaller apertures, and then corrected to the total flux of the system. 

\section{The Optical Counterparts}

We provide identifications for the low-mass stars following the IAU convention
\markcite{di87}(Dickel \etal 1987) in Table~\ref{tbl-2}. The identifications
are the truncated J2000 coordinates. We also provide CrAPMS numbers which are
shorter and easier to work with. Note that, by IAU convention, the proper
designation for CrAPMS~1 is nTT J190134-3700. In this designation, the ``nTT''
prefix is retained for continuity with other papers in this series, and does
not necessarily imply anything about the evolutionary status of the stars. 

Finding charts for newly-identified stars are presented in Figure~\ref{fig2}. A
finding chart for CrAPMS~7 is given in \markcite{mr81}Marraco \& Rydgren
(1981), and finding charts for CrAPMS~1, 2, and 3 are in \markcite{wa86}Walter
(1986). Optical spectra are shown in Figures~\ref{fig3} and \ref{fig4}. Low
dispersion spectra of CrAPMS~1, 2, and~3 are in \markcite{wa86}Walter (1986).
Comments on the individual sources follow. 

\begin{description}
\item[CrAPMS~1.] This star has many aliases: CD~-37$^o$~13022, Anon~1
              \markcite{k73}(Knacke \etal 1973), i2 \markcite{gp75}
              (Glass \& Penston 1975), VSS13 \markcite{vss} 
              (Vrba, Strom, \& Strom 1976), CrA~1 \markcite{wa86}
              (Walter 1986) and
              HBC~676 \markcite{hb88}(Herbig \& Bell 1988). \markcite{mr81}
              Marraco \& Rydgren (1981) noted
              weak Ca~II H\&K emission cores and 0.2~mag variability at V, and
              called it a PMS star with extremely weak emission.
\item[CrAPMS~2.] This is CD~-37$^o$~13029, a2 \markcite{gp75} 
              (Glass \& Penston 1975), VSS1 \markcite{vss}(Vrba \etal 1976), 
              CrA~2 \markcite{wa86}(Walter 1986),
              and HBC~678 \markcite{hb88}(Herbig \& Bell 1988).
\item[CrAPMS~3.] This is star w \markcite{gp75} (Glass \& Penston 1975),
               VSS24 \markcite{vss} (Vrba \etal 1976),
               CrA~3 \markcite{wa86} (Walter 1986),
               and HBC~679 \markcite{hb88} (Herbig \& Bell 1988). The faint
               optical companion visible to the northeast is very red, and is
               about 1.8~mag fainter than CrAPMS~3 at K (J--K~=~1.42$\pm$0.04,
               J--H~=~1.00$\pm$0.04).
\item[CrAPMS~4.] This source was found by running the MDETECT source-detection
              algorithm off-line. Two stars, designated 4NW and 4SE and
              separated by 43~arcsec, are within the X-ray error circle.
              Both are PMS stars. 
\item[CrAPMS~6.] This 3~arcsec pair of M3 stars is H$\alpha$~11 of
              \markcite{mr81} Marraco \& Rydgren. 
              The stars appear of comparable
              brightness on the Palomar Observatory Sky Survey prints and on
              our CCD images, but on 3 dates (1987 June 5, 1988 May 24, and 
              1989 May 27) the SW component appeared faint. On 1988 May 24 the
              SW component was fainter than the NE component by 2.06$\pm$0.48
              magnitudes at V. In 1993 May the V magnitudes were identical
              within the errors. 6SW may be a large amplitude variable.
              6NE is also variable, with a 0.3~mag range at V in 3
              observations. In 1997 the K~magnitudes were comparable.
\item[CrAPMS~7.] The counterpart is H$\alpha$~6 of \markcite{mr81} 
              Marraco \& Rydgren. 
              H$\alpha$~5 is on the edge of the 90\% confidence error circle,
              but it is fainter, bluer, and does not have H$\alpha$ emission
              \markcite{mr81} (Marraco \& Rydgren 1981),
              and is unlikely to be PMS.
              H$\alpha$~6 is a classical T~Tauri star, with an
              infrared excess \markcite{bw92} (Wilking 
              \etal 1992) and a strong H$\alpha$ emission line
              (W$_\lambda$(H$\alpha$)=--34\AA ).
\item[CrAPMS~9.] There are three stars in the 43~arcsec error circle.
              The star to the
              north is the PMS star; the close visual pair to the south shows
              neither evidence for H$\alpha$ emission nor Li absorption.
\item[CrAPMS~10.] This X-ray source is coincident with the pair of late-B stars
              HD~176269 (=HR7169 = SAO 210815) and
              HD~176270 (=HR7170 = SAO 210816). The resolution of the IPC is
              insufficient to split the pair.
\item[CrAPMS~11.] This source is coincident with the B9 star TY~CrA, in the
              NGC 6726/7 nebulosity. TY~CrA is a multiple system (e.g., 
              \markcite{ca93}
              Casey \etal 1993). The B9 star is part of a double-lined 
              spectroscopic binary with a 2.9$^d$ orbital period 
              \markcite{ca95} (Casey \etal
              1995). The secondary is a 1.6~M$_\odot$ G star. There is also a
              tertiary in the system.
\end{description}

We identified no spectroscopic binaries among the 6 stars observed at high
dispersion. Four of the low-mass stars are in two visual pairs. We consider
both to be physical pairs. The members of the wide CrAPMS~4 pair ($\sim$5000~AU
projected separation) have essentially identical radial velocities, and the
CrAPMS~6 pair consists of similar stars with a projected separation of 300~AU.
The spatial association of each pair seems secure, but that they are
gravitationally bound is not established, especially in the case of CrAPMS~4.
Each of the higher mass objects, CrAPMS~10 and CrAPMS~11, are multiple. The
binary fraction exceeds 56\% (9 stars of 16 in multiple systems) or 36\% (4
multiple systems of 11), depending on the definition. The binary fraction is
higher (59\% or 46\%) if we assume that the companion to CrAPMS~3 is physical.
The number of stars per system exceeds 1.4. 

\section{Discussion}
\subsection{The Low-Mass Stars}
Seven of these 11 low-mass PMS stars were first noted through their X-ray
emission, 5 here and 2 by \markcite{wa86} Walter (1986); only the H$\alpha$~11
pair (CrAPMS~6), H$\alpha$~6 (CrAPMS~7), and Anon~1 (CrAPMS~1) were first
identified using optical techniques. Once again, X-rays prove to be a powerful
search technique for low-mass PMS stars. 

The photometric magnitudes and colors are given in Tables~\ref{tbl-pht} and
\ref{tbl-irpht}. The second column of these tables gives the number of
independent optical or near-infrared observations the listed photometry is
based upon. Errors of a single observation in VRI were typically .01~mag at
V~$<$~15~mag, increasing to 0.03~mag at V~=~16~mag, while at B and U the errors
increased rapidly above 0.02 mag for V~$>$~14, due to the red nature of these
stars.   However, the uncertainties in the optical photometry listed in
Table~\ref{tbl-pht} are dominated by the fact that all stars with sufficient
numbers of observations showed evidence for variability. The range of observed
variability is typically 0.15, 0.08, 0.04, and 0.02~mag for UBVR, respectively.
For stars with many (n~$\gta$~15) observations, the variability appears
consistent with rotational modulation by starspots with additional flaring
visible in the U and B filters. In the near-IR, the uncertainties on a single
observation are typically 0.02~mag (JHK) or 0.03~mag (L). Only CrAPMS~3 showed
significant near-IR variability (Table~\ref{tbl-irpht}); part or all of this
may be attributable to its companion. 

The measured radial (V$_{rad}$) and rotational (V$sin~i)$ velocities and
equivalent widths W$_\lambda$, the estimated spectral types and extinctions
A$_V$, and the derived luminosities and masses are presented in
Table~\ref{tbl-par}. We determine the spectral types by visual comparison
of the spectra with a grid of MK standards. 
We use the spectral types and the colors jointly to estimate the
extinctions and luminosity classes. 
Temperatures are estimated from the unreddened R$_C$--I$_C$ and
V--K colors (\markcite{la85} Laird 1985, \markcite{ca83} Carney 1983,
\markcite{ve74} Veeder 1974), and agree with the spectral type temperature
calibration of \markcite{jn87} de Jager \& Nieuhenhuijzen (1987), with a mean
difference of 39K. For stars with V--I$_C>$1.57, we use the color-T$_{eff}$
calibration of \markcite{k93}Kirkpatrick \etal (1993).
\markcite{st97}Stauffer (1997) has shown that this color-T$_{eff}$ calibration
calibration brings the lower main sequence of the Pleiades into agreement
with the \markcite{dm94} D'Antona \& Mazzitelli (1994) evolutionary tracks.
Masses and ages are determined from the
star's location on the H-R diagram, and are dependent upon the choice of
evolutionary track. Further details of the techniques may be found in
\markcite{wa94} Walter \etal (1994). We assume a distance of 130~pc to the
association. On the basis of spatial association with the R~CrA cloud, location
above the main-sequence, luminosity classes, radial velocities, and the
presence of Li~I~$\lambda$6707\AA\ absorption, all these stars are PMS. The
dispersion in radial velocities is consistent with the 2~km~s$^{-1}$ velocity
dispersion generally observed among PMS stars in the central parts of
star-forming regions \markcite{m86} (Mathieu 1986). 

The lithium abundances log~N(Li) (Table~\ref{tbl-par}) are derived using the
NLTE analysis of \markcite{pm96} Pavlenko \& Magazz\`u (1996). These curves of
growth are valid for T$_{eff}>$3500K; we extrapolate their curves of growth for
our coolest stars (3300K$<$T$_{eff}<$3500K). The abundances determined for the
hotter stars are consistent with the abundances seen in T~Tauri stars
\markcite{bmb91} (Basri, Martin, \& Bertout 1991). These are presumably the
initial Li abundances, as depletion is not yet expected in stars of these
masses. For the cooler stars, the abundances are seen to be significantly
lower. Pre-main sequence lithium burning (e.g., \markcite{pkd90} Pinsonneault,
Kawaler, \& Demarque 1990) is expected among the M stars, but the expected
depletion of less than half a dex for an age of $<$10~Myr is significantly less
than we measure. This discrepancy may be accounted for by the fact that the
presence of an active chromosphere/transition region will over-ionize Li,
giving an apparent underabundance \markcite{hd95} (Houdebine \& Doyle 1995).
Note that the M stars with the largest H$\alpha$~equivalent widths have the
smallest apparent Li abundances. 

There is evidence of a near-IR excess attributable to warm circumstellar
dust only in H$\alpha$~6 (CrAPMS~7).

X-ray count rates and derived luminosities, surface fluxes and flux ratios are
given in Table~\ref{tbl-xray}. None of the targets were sufficiently bright to
make extraction of the spectrum practical. We assume that the X-ray emission
arises in a 10$^7$K corona for conversion to luminosities and flux ratios. All
derived quantities in Table~\ref{tbl-xray} are corrected for extinction,
assuming R=3.1. We divide the observed counts equally between the stars in the
two pairs. This is a good assumption for CrAPMS~6, which consists of two
similar stars, but probably not for the CrAPMS~4 pair. The luminosities and
flux ratios are similar to those observed in other young associations
\markcite{wa94}(Walter \etal 1994). 

The luminosity-temperature diagram for the X-ray-selected stars is presented in
Figure~\ref{fig5}. The two panels show the stars plotted against theoretical
evolutionary tracks from \markcite{dm94}D'Antona \& Mazzitelli (1994), and
from \markcite{sw94}Swenson \etal (1994). The D'Antona \& Mazzitelli tracks
employ the \markcite{cm90}Canuto \& Mazzitelli (1990) treatment of convection
and \markcite{a89}Alexander \etal (1989) opacities. Both sets of tracks are
for solar abundances.

In the absence of other information, we must rely on the
evolutionary tracks to estimate stellar masses and ages.
We note the following in Figure~\ref{fig5}:
\begin{enumerate}
\item The X-ray-selected stars are significantly older than the IR-excess
    stars discussed by \markcite{k73}Knacke \etal
    (1973) and by \markcite{bw92}Wilking \etal (1992). Note that the X-ray
    and IR samples have few stars in common.
    Ages estimated from the Swenson \etal tracks are systematically older than
    those from the D'Antona \& Mazzitelli tracks, by about a factor of 2.
    The mean ages are 7 and 15~Myr, relative to the D'Antona \& Mazzitelli
    and the Swenson \etal tracks, respectively.
\item With the exception of CrAPMS~1, which appears significantly younger
    than the rest of the stars, the stars lie along a narrow locus.
\item There is no evidence that the H$\alpha$-emitting stars H$\alpha$~6 and
    H$\alpha$~11 are significantly younger than the other X-ray-selected stars.
\end{enumerate}

\subsection{X-ray Emission from Higher Mass Stars}
Two of the X-ray sources are coincident with late-B stars. TY~Cra (B9V) is an
eclipsing triple system \markcite{ca95} (Casey \etal 1995). With a mass of
1.6~M$_\odot$ and an age of $<$6~Myr, the secondary will be a G/K star similar
to HDE283572 \markcite{wa87} (Walter \etal 1987), and can easily account for
all the observed X-ray flux. The tertiary mass is less certain: \markcite{ca95}
Casey \etal (1995) find 2.4~M$_\odot$, while \markcite{clb96} Corporon,
Lagrange, \& Beust (1996) find a mass of 1.2-1.4~M$_\odot$. In either case the
tertiary is likely to be an X-ray source. Both HR7169 (B9V) and HR7170
(B8IV-V), a 13~arcsec visual pair,
are listed as spectroscopic binaries in the Yale Bright Star catalog 
\markcite{ybs} (Hoffleit 1982); further observations are needed to determine
if there are low-mass stars associated with the HR7169/70 system. 

Are late-B stars in star-forming regions bright X-ray sources? A basic
conclusion to be drawn from the EINSTEIN and ROSAT stellar X-ray observations
is that all stars are X-ray sources --- {\it except the late-B to mid-A stars}.
\markcite{js93} Schmitt \etal (1993) observed the young \markcite{l86} Lindroos
(1986) pairs, consisting of a near-zero-age main sequence B star and a young
late-type visual companion. Using the ROSAT HRI, with 3~arcsec positional
accuracy, they showed that in 4 cases {\it both} the B star and the late-type
companion (ages less than 5$\times$10$^7$ years) were X-ray sources. However,
\markcite{cgs94} Caillault, Gagn\'e, \& Stauffer (1994) have argued that the
fraction of late-B/early-A stars in the Orion Nebula which are detected as
X-ray sources is just that fraction expected to have late-type (F5 or later)
binary companions. Our data clearly do not help resolve the question of whether
the late-B stars can be X-ray sources. 

\subsection{X-ray Emission from Classical T~Tauri Stars}
The classical T~Tauri stars S~CrA and DG~CrA, and the intermediate-mass stars
R~CrA and T~CrA were within the area viewed by the EINSTEIN observatory. None
were detected. The 2$\sigma$ upper limit to the IPC count rates is given in
Table~\ref{tbl-3}. The conversion from flux to counts depends on the assumed
extinctions, which are poorly known. We assumed a coronal plasma at a
temperature of 10$^7$K. For a wide range of reasonable extinctions
(A$_V\lta$~10~mag), the count rate limits correspond to coronal luminosities
L$_X<$5$\times$10$^{29}$~erg~s$^{-1}$. 

\subsection{The Population of the R~CrA Associations}

Star formation in this cloud has been going on for a long time.
The X-ray-selected stars, with ages extending upward to over 10~Myr, appear
significantly older than the B stars and imbedded sources. The narrow locus
occupied by the stars in the luminosity-temperature diagram may be evidence
for a burst of star formation about 7-10~Myr ago, since younger stars
will be more luminous, both at optical and X-ray wavelengths (that the imbedded
IR sources are not detected is a consequence of their high extinction and the
relatively high limiting flux in these short observations; 
\markcite{K96}Koyama \etal (1996) and \markcite{rn97}Neuh\"auser \& Preibisch
(1997), using ASCA and ROSAT, respectively, have reported the X-ray detection
of imbedded class~I protostars). However, given the small numbers, we cannot
rule out continuous star formation at a constant rate over the last 10~Myr.
Given this
population of older stars, it appears that the R~CrA cloud is not simply a
lower mass analog of the $\rho$~Oph cloud \markcite{bw92}(Wilking \etal 1992),
but may represent a more evolved state of such a cloud. 

We observe 11 low-mass stars (counting visual pairs as 2 stars, and neglecting
the probable low-mass companions of the B stars) in these 9 X-ray error
circles. In previous studies of the X-ray emitting populations of star-forming
associations \markcite{wa88}\markcite{wa94} (Walter \etal 1988; 1994), we
argued that the completeness of the X-ray sample is $\sim$30\%. This
incompleteness has been verified in the Taurus region by \markcite{n95}
Neuh\"auser \etal (1995). The same arguments hold here, implying that the total
number of X-ray emitting low-mass PMS stars in the $\sim$1.5~square-degrees
observed is about 30. Therefore the total number of low-mass stars in this
association must be about 60, or double that previously known. 


This is an underestimate of the total population of the R~CrA association. The
X-ray observations sampled only about 1.5~square degrees near the center of the
cloud. Over 10\% (2 of 16) of those IR sources whose spectral energy
distributions are consistent with their being PMS stars \markcite{bw92}
(Wilking \etal 1992 -- see Figure~\ref{fig1}) lie in the dark streamers to the
east of the R~CrA complex, near the B~cloud \markcite{R78}(Rossano 1978) and
outside the EINSTEIN IPC fields. If the X-ray
sources are spatially distributed in the same way, the total number of X-ray
emitting low-mass PMS stars must be increased by over 10\%. Note that
\markcite{ne97}Neuh\"auser \etal (1997) identified
GSC~7916~0050, 1.6$^o$ to the southwest of R~CrA and well outside the
IPC fields, as a PMS star. In addition, some
fraction of the X-ray bright population will lie behind the dark cloud, and
will not be detected because their X-ray fluxes lie below our flux limit. Since
4 PMS systems are projected against the near side of the cloud, we would expect
about the same number, or about 30\% of the total population, to be hidden from
view. These corrections would increase the inferred X-ray bright population
(seen and unseen) to about 40 stars, and the total population to about 70
stars. 

\markcite{bw94}Wilking \etal (1994) claimed, based on IR
source counts, that the total stellar population in the core 0.24~pc$^2$
of the CrA complex is small
($\sim$30 stars), assuming a 2~Myr age for the association. If the age is
10~Myr, their IR counts permit up to 78 association members in this small
region (Wilking, private communication). 
Given the large uncertainties in the estimates and the differences in the
volumes samples, there is no significant difference in the total population
estimates from the IR or X-ray samples.

The X-ray-bright population is observed to be more widely dispersed in space
than the IR-selected population. This is to be expected if the X-ray-bright
population is older, and has some non-zero velocity dispersion. Some fraction
of the total population may have dispersed outside the area we observed. To
estimate the fraction of an old, dispersed population that should be visible
within the volume sampled by the EINSTEIN observations, we modelled the
association as a population with an age of 7~Myr and a velocity dispersion of
1.7~km~s$^{-1}$, and neglect gravitational decelerations.
We assume that the stars all form within 1$^o$ of the position
of R~CrA. If the star formation rate has remained constant over the past 7~Myr,
then about 70\% of the members of the association will lie within the volume
sampled. If the X-ray bright sample represents a short burst of star formation,
with an age spread of 1~Myr, then only 45\% of the population lies within the
area sampled. Identifying the IR-bright sample with the youngest sources, and
the X-ray bright sample with the non-imbedded older population, then the
correction for the stars outside the IPC fields leads to an estimate of a total
population of about 90 stars (in the case of continuous star formation,
N=70$\times$1.3; in the case of 2 episodes of star formation,
N=30+40$\times$1.55), with a likely uncertainty of $\pm$20 stars. 

We are
attempting to verify this estimate using an X-ray-selected sample of stars from
observations with the ROSAT Observatory (Walter \etal, in preparation). 

\markcite{bw92}Wilking \etal (1992) noted that 25\% of the stars in this
association are of intermediate-mass (spectral types B8 to A5),
an unusually large
number. Other star formation regions (e.g., Upper Sco; \markcite{wa94}Walter
\etal 1994) show initial mass functions (IMF) similar to the field star mass
function. If the R~CrA IMF is consistent with the \markcite{ms79}Miller-Scalo
(1979) field star mass function, and the intermediate mass stars all have
masses greater than 2.5$\pm$0.5~M$_\odot$, then one expects a total population
of 92$^{+34}_{-27}$ stars with M$>$0.2M$_\odot$. Given the uncertainties in
both the extrapolation of the IMF from 6 high mass stars and in our estimate of
the low mass population, we see no evidence that the the mass function is
abnormal. We conclude that the CrA clouds have spawned of order 100 low-mass
stars over the past 10~Myr, and that there is no compelling evidence for an
excess of intermediate-mass stars. 

\section{Summary}

We have identified 11 low-mass pre-main-sequence X-ray emitting stars
associated with the R~CrA star-forming region. With the exception of
H$\alpha$~6, none of these stars have near-IR excesses attributable to warm
circumstellar dust. None of the four classical T~Tauri stars in the field were
detected. X-ray emission was detected from the vicinity of the late-B stars
TY~CrA and HR~7169, but the emission may be attributable to lower mass
companions. 

Only 4 of the low-mass stars (2 in a close visual pair) were previously known
from optical work to be PMS. The low-mass stars are significantly older than
the 1-2~Myr age inferred for the infrared sources and intermediate-mass stars.
The narrowness of the locus occupied by the X-ray-selected stars suggests an
episode of star formation some 7~Myr ago, but we cannot exclude continuous
star formation at a roughly constant rate over the past 10~Myr.

We estimate that the older nTT population dominates the total number of stars,
and that the total PMS population amounts to about 90 stars.
Given 6 intermediate-mass stars, the number of low-mass stars is consistent
(within the large uncertainties) with the field initial mass function.

\acknowledgements

This research has been supported by NSF grant AST89-96308, NASA grants
NAG8-884, NAG5-1663 (ADP), and NAGW-1365 (Origins of Solar Systems)
to SUNY, and NSF grant AST8814986 to the University of Wisconsin. RDM
acknowledges support from the Presidential Young Investigator Program.
The paper is based in part on observations collected at the European
Southern Observatory, La Silla, Chile.
We thank the folks at CfA, especially
F.~Seward and J.~McSweeney, for their assistance with obtaining and examining
the IPC images and using the measuring engine. 
We thank R.~Millis for contributing
a portion of the UBV photometry, and J.~Stauffer for bringing to our attention
an improved color--T$_{eff}$ relation for M stars.
We thank R.~Davis, E.~Horine, J.~Peters,
D.~Latham, and A.~Milone for assistance in the acquisition of the CfA radial
velocity measurements, and Juan Alcal\'a of MPE Garching
and the ESO staff at La Silla Observatory.
This research made use of the SIMBAD database,
operated by CDS, Strasbourg, France, and of the Digitized Sky Survey, produced
at the Space Telescope Science Institute under grant NAG W-2166. We thank the
referee for constructive criticisms and for bringing to our attention the
near-IR brightness of the faint optical companion to CraPMS~3.

\begin{table}
\dummytable\label{tbl-1}
\end{table}

\begin{table}
\dummytable\label{tbl-2}
\end{table}

\begin{table}
\dummytable\label{tbl-pht}
\end{table}

\begin{table}
\dummytable\label{tbl-irpht}
\end{table}

\begin{table}
\dummytable\label{tbl-par}
\end{table}

\begin{table}
\dummytable\label{tbl-xray}
\end{table}

\begin{table}
\dummytable\label{tbl-3}
\end{table}

\newpage

\figcaption[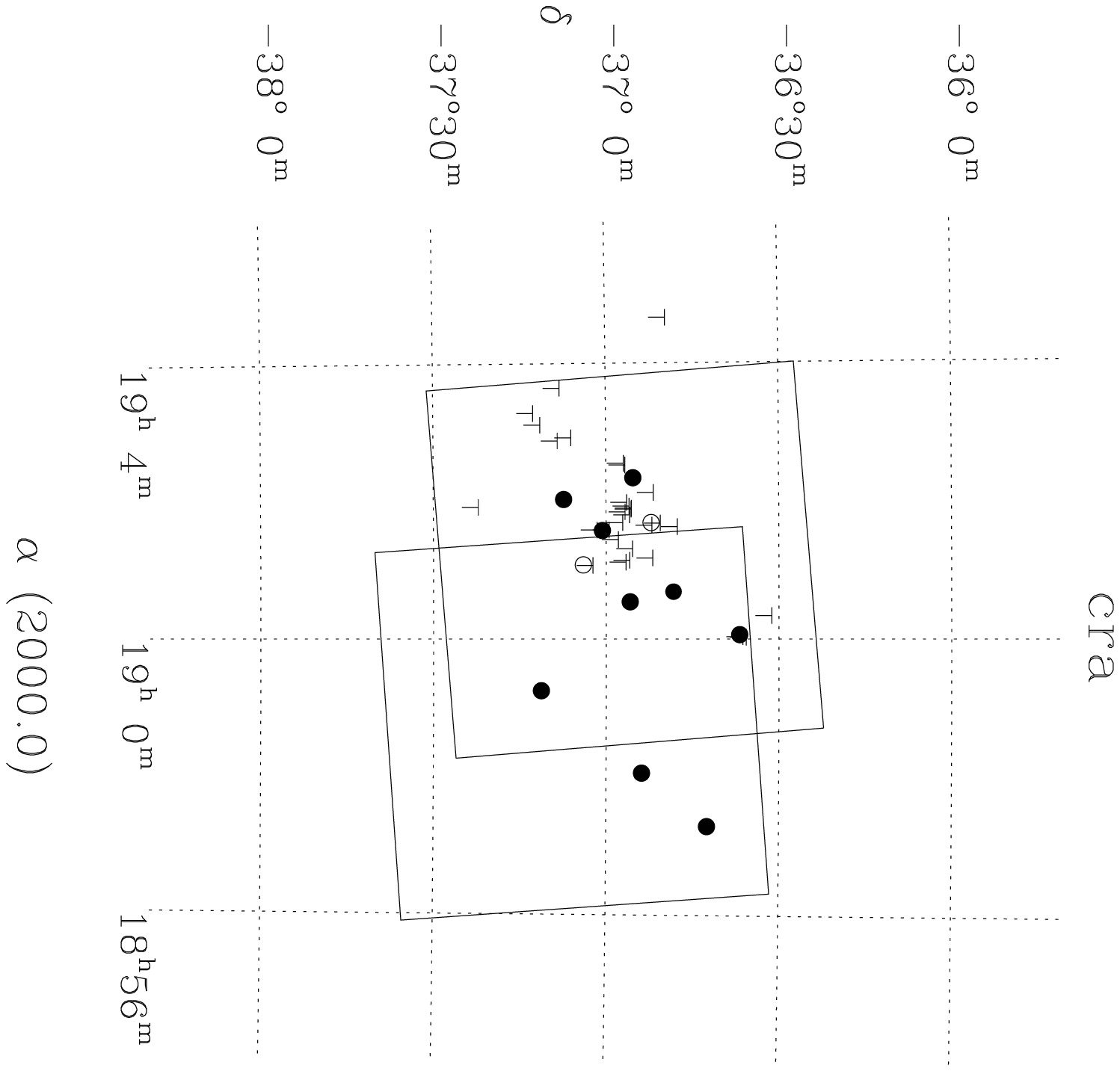]{
          EINSTEIN Fields. The locations of the low-mass PMS stars are
          indicated by the filled circles. Open circles mark the B
          star systems TY~CrA and HR~7169/70.
          Near-IR sources from 
          compilation are marked by the `T's. \label{fig1}}

\figcaption[walter.fig2.eps]{
          Finding charts for the 6 newly-identified PMS stars.  
          Each image is 4~arcmin on a side.  These images are taken from the 
          Digitized Sky Survey and have 1.7 arcsecond pixels.  North
          is up and east is to the left.  \label{fig2}}

%

\figcaption[walter.fig3a.ps]{
          Low-dispersion spectra. The data have been smoothed with a Fourier
          filter.  The Y axis scale runs from zero to 120\% of the maximum flux
          in the plotted interval. The prominent emission lines in the left 
          panels are the chromospheric Ca~II K\&H lines. H$\alpha$ is the
          prominent emission line in the red spectra.
    \label{fig3}}

\figcaption[walter.fig4.ps]{
          High-dispersion spectra. The discontinuities sometimes visible near 
          6465\AA\ and 6535\AA\ are the ends of the echelle orders. Plot scaling
          and smoothing are as in Figure~3. The left panel shows the H$\alpha$
          emission line and a 150\AA\ stretch of continuum. A wide
          variety of H$\alpha$ line profiles is evident. The right panel
          shows the Li~I ($\lambda$6707\AA ) and Ca~I ($\lambda$6717\AA )
          lines. Rotational broadening is evident in some spectra.
   \label{fig4}}

\figcaption[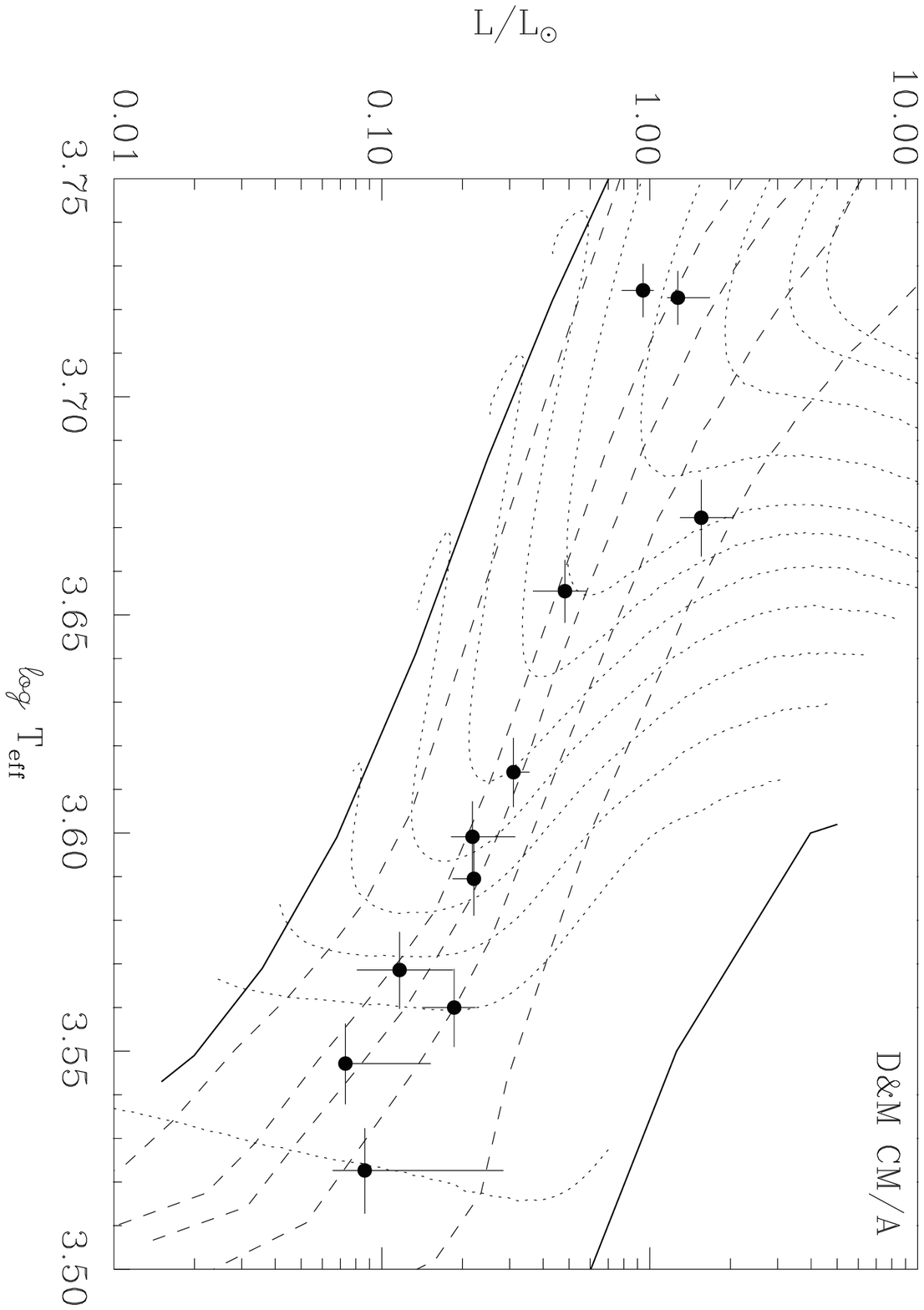]{
          Two luminosity-temperature diagrams: one using the 
          D'Antona \&
          Mazzitelli (1994) evolutionary tracks; the other using the 
          Swenson \etal (1994) tracks. Isochrones (dashed) are
          at ages of 1, 3, 6, 10, and 30~Myr;
          evolutionary tracks (dotted) are plotted for
          0.2, 0.4, 0.5, 0.6, 0.7, 0.8, 0.9, 1.0, 1.2, 1.5, 1.8, and 2.0
          M$_\odot$. The birthline 
          (Stahler 1988) and zero-age main sequence 
          (Pinsonneault \etal 1990) are overplotted as solid lines
          for reference. Uncertainties in luminosity mainly indicate the
          uncertainty in the extinction correction.
\label{fig5}}
\clearpage
\begin{deluxetable}{crrcl}
\tablenum{1}
\tablecaption{EINSTEIN IPC POINTINGS}
\tablehead{
\colhead{Image} & \multicolumn{2}{c}{$\alpha$ (B1950) $\delta$} &
\colhead{Time (sec)} & \colhead{Target} \\
                & \multicolumn{2}{c}{(decimal degrees)} }

\dummytable\label{tbl-1}


\startdata
3501 & 283.849 & -37.167 & 2128 & $\epsilon$~CrA \\
4512 & 284.445 & -37.016 & 2427 & S~CrA \\
\enddata
\end{deluxetable}
\begin{deluxetable}{llrrll}
\tablenum{2}
\dummytable\label{tbl-2}
\tablecaption{Positions of the CrA Stellar X-Ray Sources}
\tablehead{
\colhead{nTT J} & \colhead{CrAPMS} & \colhead{GSC}
  & \multicolumn{2}{c}{$\alpha$ (J2000) $\delta$} & \colhead{Other}}

\startdata
185717-3643 &4NW & 7421~0665 & 18 57 17.8  & -36 42 35.9 & \\
185720-3643 &4SE & 7421~1242 & 18 57 20.7  & -36 43 00.4 & \\
185801-3655 &5   & 7421~1040 & 18 58 01.7  & -36 53 45.0 & \\
185914-3711NE\tablenotemark{a} &6NE &    & 18 59 14.7  & -37 11 30 &H$\alpha$~11\\
185914-3711SW &6SW  & & \tablenotemark{b} & \tablenotemark{b} & H$\alpha$~11\\ 
190001-3637\tablenotemark{c} &7  &  & 19 00 01.5  & -36 37 07 & H$\alpha$~6 \\
190028-3656\tablenotemark{a} &8  &  & 19 00 28.9  & -36 56 02 & \\ 
190039-3648\tablenotemark{a} &9  &  & 19 00 39.1  & -36 48 11 & \\ 
\multicolumn{1}{c}{---}&10&7421~2294 & 19 01 03.1 & -37 03 38.7 & HR7169\\
\multicolumn{1}{c}{---}&~"~&7421~2295 &19 01 04.2 & -37 03 41.8 & HR7170\\
190134-3700 &1   & 7421~1890 & 19 01 34.9  & -37 00 55.8 & CrA~1\tablenotemark{d}\\
\multicolumn{1}{c}{---}&11&7421~1126 & 19 01 40.8 & -36 52 36.3 & TY CrA\\
190201-3707 &2   & 7421~0493 & 19 02 01.9  & -37 07 43.2 & CrA 2\tablenotemark{d}\\
190222-3655 &3   & 7421~1899 & 19 02 22.1  & -36 55 40.8 & CrA 3\tablenotemark{d}\\

\tablenotetext{a}{Coordinates from Digitized Sky Survey, and represent the
      centroid of the
      light. These coordinates are not in the Guide Star frame.}
\tablenotetext{b}{The SW component is 0.22$^s$ west and 1.7'' south of the NE component.}
\tablenotetext{c}{Coordinates from Marraco \& Rydgren 1981.}
\tablenotetext{d}{Walter 1986.}

\enddata
\end{deluxetable}
\begin{deluxetable}{lrrrrrrr}
\tablenum{3}
\dummytable\label{tbl-pht}
\tablecaption{Visual Photometric Data}
\tablehead{
\colhead{CrAPMS} & \colhead{n} & \colhead{V} & \colhead{$\Delta$V}& 
\colhead{U-B}  & \colhead{B-V}  & \colhead{V-R$_C$}  & \colhead{R$_C$-I$_C$} }
\startdata
1       & 20 & 11.34 & 0.21 & 0.65 & 1.11 & 0.71 & 0.73 \\
2       & 18 & 10.44 & 0.15 & 0.29 & 0.79 & 0.47 & 0.46 \\
3       & 18 & 13.79 & 0.18 & 1.04 & 1.43 & 0.99 & 1.02 \\
4NW     &  5 & 13.35 & 0.02 & 1.16 & 1.38 & 0.93 & 0.96 \\
4SE     & 13 & 10.95 & 0.07 & 0.44 & 0.88 & 0.51 & 0.49 \\
5       & 15 & 12.30 & 0.20 & 0.85 & 1.11 & 0.70 & 0.67 \\
6NE     &  3 & 15.14 & 0.36 & 1.03 & 1.48 & 1.11 & 1.56 \\
6SW\tablenotemark{a} &1&15.34 &\nodata&\nodata & 0.96 & 1.26 & 1.82  \\
7\tablenotemark{b}   &1&13.93 &\nodata&\nodata & 1.26 & 0.85 & 1.00  \\
8       &  3 & 15.68 & 0.14 & 1.20 & 1.47 & 1.30 & 1.66 \\
9       &  1 & 15.88 & \nodata & 1.47 & 1.65 & 1.26 & 1.58 \\

\tablenotetext{a}{ The observed V magnitude was 17.14$\pm$0.48 on 1988 May 24.}
\tablenotetext{b}{from \markcite{mr81}
      Marraco \& Rydgren (1981), and converted to the
      Cousins system using the transformation of \markcite{kr79}
      Kunkel \& Rydgren (1979).}

\enddata
\end{deluxetable}
\begin{deluxetable}{lrrrrr}
\tablenum{4}
\dummytable\label{tbl-irpht}
\tablecaption{Near-IR Photometric Data}
\tablehead{
\colhead{CrAPMS}& \colhead{n} & \colhead{K\tablenotemark{a}}& 
\colhead{J-K}  & \colhead{J-H}  & \colhead{K-L}}
\startdata
1                 & 4 & 7.80 & 0.92 & 0.68 & 0.13 \\
2                 & 3 & 8.34 & 0.52 & 0.43 & 0.08 \\
3\tablenotemark{c}&2& 9.18 & 1.06 & 0.80 & 0.16 \\
3\tablenotemark{d}&2& 9.05 & 1.11 & 0.83 &\nodata \\
3\tablenotemark{e}&1& 9.21 & 1.12 & 0.84 &\nodata \\
4NW               & 2 & 9.42 & 0.84 & 0.67 &\nodata  \\
4SE               & 2 & 8.70 & 0.57 & 0.47 &\nodata  \\
5                 & 4 & 9.20 & 0.73 & 0.59 & 0.11 \\
6NE               & 1 &10.69 & 0.77 & 0.54 &\nodata  \\
6NE               & 1 &10.56 & 0.84 & 0.55 &\nodata  \\
7\tablenotemark{b}&- & 9.35 & 1.43 & 0.93 & 0.66 \\
8                 & 2 & 9.59 & 1.12 & 0.81 &\nodata  \\

\tablenotetext{a}{The standard deviation of the mean is 0.02~mag
     for single observations, and 0.01~mag otherwise.
     Uncertainty on the colors is $\pm$0.04~mag for
     single observations, and $\pm$0.02~mag otherwise.}
\tablenotetext{b}{ Data from Wilking {\it et al.\/} (1992).}
\tablenotetext{c}{ Observations on 1986 June 2 and 3.}
\tablenotetext{d}{ Observations on 1989 May 26 and 27.}
\tablenotetext{e}{ Observations on 1997 February 21.}
\enddata
\end{deluxetable}
\begin{deluxetable}{lllrrrlrrrrrr}
\tablenum{5}
\dummytable\label{tbl-par}
\tablecaption{Stellar Parameters}
\tablehead{
\colhead{CrAPMS}   & \colhead{Type} & \colhead{A$_V$\tablenotemark{a}}&
\colhead{V$_{rad}$} &
\colhead{V$_{rot}$}& \colhead{n\tablenotemark{b}} &
\colhead{W$_\lambda$(Li)} &
\colhead{log}                       & 
\colhead{W$_\lambda$(H$\alpha$)} & 
\colhead{T$_{eff}$} &
\colhead{log} & 
\colhead{M/M$_\odot$\tablenotemark{d}}\\
&&&&&  &  &\colhead{N(Li)\tablenotemark{c}} & & & \colhead{(L/L$_\odot$)}}

\startdata
1   & K1IV  &1.2 & -1.0  &58     & 8& 0.39  & 3.0   &   0.3& 4702&  0.19 &1.0 \\
2   & G5IV  &0.3 & -1.2  &23     &20& 0.28  & 3.1   &  -0.7& 5281&  0.10 &1.2 \\
3   & K2IV  &2.3 & -1.2  &$<$10  & 4& 0.41  & 3.2   &  -0.9& 4523& -0.31 &0.9 \\
4NW & M0.5V &0.2 & -2.2  &$<$10  & 5& 0.45  & 2.0   &  -1.1& 3886& -0.66 &0.6 \\
4SE & G5IV  &0.5 & -2.0  &$<$10  & 6& 0.36: & 3.5   &   1.0& 5302& -0.03 &1.1 \\
5   & K5IV  &0.1 & -0.8  &22     & 7& 0.44  & 2.4   &  -0.8& 4111& -0.51 &0.7 \\
6NE & M3V   &0.2 &\nodata&\nodata& -& 0.70  & 2.2   &  -5.9& 3525& -1.14 &0.3 \\
6SW & M3.5V &0.3 &\nodata&\nodata& -& 0.44  & 1.7   &  -9.8& 3332& -1.06 &0.2 \\
7   & M1V   &0.6 &\nodata&\nodata& -& 0.36: & 1.5   & -33.5& 3973& -0.66 &0.7 \\
8   & M3V   &1.6 &\nodata&\nodata& -& 0.57  & 2.1   &  -3.9& 3631& -0.73 &0.4 \\
9   & M2V   &1.7 &\nodata&\nodata& -& 0.5:  & 1.9   &  -9.2& 3704& -0.93 &0.5 \\

\tablenotetext{a}{Uncertainties typically $\pm$0.2~mag.}
\tablenotetext{b}{Number of velocity observations.}
\tablenotetext{c}{Logarithmic Li abundances, interpolated from a grid by
                  Pavlenko \& Magazz\`u (1996) for T$_{eff}>$3500K, 
                  and extrapolated for T$_{eff}<$3500K. Uncertainties based on
                  low dispersion spectra (equivalent widths followed by ``:'')
                  are roughly $\pm$0.3~dex. }
\tablenotetext{d}{Masses estimated from D'Antona \& Mazzitelli (1994) tracks.}

\enddata
\end{deluxetable}
\begin{deluxetable}{lrccr}
\tablenum{6}
\dummytable\label{tbl-xray}
\tablecaption{X-Ray Fluxes }
\tablehead{
\colhead{CrAPMS} & \colhead{c s$^{-1}$\tablenotemark{a}} & 
\colhead{L$_X$\tablenotemark{b}} & \colhead{F$_X$} &
\colhead{$\frac{L_X}{L_{bol}}$}\\
 & & \colhead{ergs s$^{-1}$} & \colhead{ergs~cm$^{-2}$~s$^{-1}$ }}
\startdata
1                    & 0.034 $\pm$0.005 & 2.9E30 & 1.4E7 & 4.5E-4\\
2                    & 0.038 $\pm$0.006 & 2.2E30 & 1.8E7 & 4.0E-4\\
3                    & 0.012 $\pm$0.003 & 1.7E30 & 1.5E7 & 7.2E-4\\
4NW\tablenotemark{c} & 0.014 $\pm$0.005 & 3.9E29 & 3.7E6 & 3.7E-4\\
4SE                  &                  & 4.6E29 & 5.1E6 & 1.1E-4\\
5                    & 0.014 $\pm$0.005 & 6.3E29 & 8.9E6 & 5.2E-4\\
6NE\tablenotemark{c} & 0.011 $\pm$0.004 & 2.9E29 & 5.2E6 & 7.1E-4\\   
6SW                  &                  & 3.1E29 & 3.7E6 & 7.3E-4\\   
7                    & 0.017 $\pm$0.004 & 1.1E30 & 2.9E7 & 1.1E-3\\
8                    & 0.010 $\pm$0.003 & 9.6E29 & 9.2E6 & 1.1E-3\\
9                    & 0.009 $\pm$0.003 & 8.9E29 & 1.4E7 & 1.5E-3\\

\tablenotetext{a}{Observed count rate, broad band.}
\tablenotetext{b}{assuming emission from an equilibrium Raymond-Smith
                  plasma at log~T=7.0.}
\tablenotetext{c}{Counts are divided equally between the two members of the
                  pair.}

\enddata
\end{deluxetable}
\begin{deluxetable}{lc}
\tablenum{7}
\dummytable\label{tbl-3}
\tablecaption{Upper Limits On Other YSOs}
\tablehead{
\multicolumn{1}{c}{YSO} & \colhead{IPC Count Rate}\\
                        & \colhead{c~s$^{-1}$ (2$\sigma$)}}
\startdata
R Cra & $<$0.004 \\
S CrA & $<$0.004 \\
T CrA & $<$0.003 \\
DG CrA &$<$0.006
\enddata
\end{deluxetable}

\newpage
\begin{references}


\reference{a89} Alexander, D.R., Augason, G.C., \& Johnson, H.R. 1989, 
     ApJ, 345, 1014

\reference{bh82} Barnes, J.V. \& Hayes, D.S. 1982,
     IRS Standard Star Manual (AURA)

\reference{bmb91} Basri, G., Martin, E.L., \& Bertout, C. 1991, A\&A, 252, 625

\reference{cgs94} Caillault, J.-P., Gagn\'e, M., \& Stauffer, J.R. 1994,
     ApJ, 432, 386

\reference{cm90} Canuto, V.M., and Mazzitelli, I. 1990, ApJ, 370, 295

\reference{ca83} Carney, B.W. 1983. AJ, 88, 623

\reference{ca93} Casey, B.W., Mathieu, R.D., Suntzeff, N.B., Lee, C.,
     \& Cardelli, J.A. 1993, AJ, 105, 2276

\reference{ca95} Casey, B.W., Mathieu, R.D., Suntzeff, N.B., \& Walter, F.M.
     1995, AJ, 109, 2156

\reference{clb96} Corporon, P., Lagrange, A.M., \& Beust, H. 1996, A\&A,
     310, 228

\reference{c80} Cousins, A.W.J. 1980, S.Afr.Astron.Obs.Circ., 1, 234

\reference{cd94} Cruddace, R.G. \& Dupree A.K. 1984, ApJ, 277, 263

\reference{dm94} D'Antona, F. \& Mazzitelli, I. 1994, ApJS, 90, 467

\reference{d87} Dame, T.M., \etal 1987, ApJ, 322, 706

\reference{jn87} de Jager, C. \& Nieuwenhuijzen, H. 1987, A\&A, 177, 217

\reference{di87} Dickel, H.R., Loret, M.-C., \& de Boer, K.S. 1987, A\&AS, 68,
     75


\reference{e82} Elias, J.H., Frogel, J.A., Matthews, K., \& Neugebauer, G.
     1982, AJ, 87, 1029

\reference{fk81} Feigelson, E.D. \& Kriss, G.A. 1981, ApJL, 248, L35

\reference{f93} Feigelson, E.D., Casanova, S., Montmerle, T., \& Guibert, J.
     1993, ApJ, 416, 623

\reference{g79} Giacconi, R. \etal 1979, ApJ, 230, 540

\reference{gp75} Glass, I.S. \& Penston, M.V. 1975, MNRAS, 172, 227


\reference{hb88} Herbig, G.H. \& Bell, K.R. 1988, Third Catalog of
     Emission-Line Stars of the Orion Population, Lick Obs. Bull. No.~1111

\reference{ybs} Hoffleit, D. 1982, {\it The Bright Star Catalogue},
   4$^{th}$ edition. (Yale University Observatory: New Haven)

\reference{hd95} Houdebine, E.R. \& Doyle, J.G. 1995, A\&A, 302, 861

\reference{j63} Johnson, H.L. 1963, Basic Astronomical Data, edited by K. Aa.
     Strand, (Univ. of Chicago Press, Chicago), p.~204

\reference{k93} Kirkpatrick, J.D., Kelley, D.M., Rieke, G.H., Liebert, J., 
     Allard, F., \& Wehrse, R. 1993, ApJ, 402, 643

\reference{k73} Knacke, R.F., Strom, K.M., Strom, S.E., Young, E.,
     \& Kunkel, W. 1973, ApJ, 179, 847

\reference{k96} Koyama, K., Hamaguchi, K., Ueno, S., Kobayashi, N.,
     \& Feigelson, E.D. 1996, PASJ, 48, L87

\reference{kr79} Kunkel, W.E. \& Rydgren, A.E. 1979, AJ, 84, 633

\reference{la85} Laird, J.B. 1985, ApJS, 57, 389

\reference{al83} Landolt, A.U. 1983, AJ, 88, 439

\reference{al92} Landolt, A.U. 1992, AJ, 104, 340

\reference{l92} Latham, D.W. 1992, in Complementary Approaches to Double and
                Multiple Star Research, IAU Colloquium No.~135, ASP Conf.
                Series 32, edited by H.A. McAlister \& W.I. Hartkopf, p. 110

\reference{l86} Lindroos, K.P. 1986, A\&A, 156, 233

\reference{l79} Loren, R.B. 1979, ApJ, 227, 832

\reference{m86} Mathieu, R.D. 1986, in Highlights of Astronomy, edited
     by J.-P. Swings, p. 481

\reference{mr81} Marraco, H.G. \& Rydgren, A.E. 1981, AJ, 86, 62

\reference{ms79} Miller, G.E. \& Scalo, J.M. 1979, ApJS, 41, 513

\reference{m83} Montmerle, T., Koch-Miramond, L., Falgarone, E.,
        \& Grindlay, J.E. 1983, ApJ, 269, 182

\reference{mu83} Mundt, R., Walter, F.M., Feigelson, E.D., Finkenzeller, U.,
     Herbig, G.H., \& Odell, A.P. 1983, ApJ, 269, 229

\reference{n95} Neuh\"auser, R., Sterzik, M.F., Schmitt, J.H.M.M.,
     Wichmann, R., \& Krautter, J. 1995, A\&A, 295, L5

\reference{ne97} Neuh\"auser, R., Thomas, H.-C., Danner, R., Peschke, S., \&
     Walter, F.M. 1997, A\&A, 318, L43

\reference{rn97} Neuh\"auser, R. \& Preibisch, T. 1997, A\&A, 322, L37

\reference{n94} Nordstrom, B., Latham, D.W., Morse, J.A., Milone, A.A.E.,
   Kurucz, R.L., Andersen, J., \& Stefanik, R.P. 1994, A\&A, 287, 338

\reference{pm96} Pavlenko, Y.V. \& Magazz\`u, A. 1996, A\&A, 311, 961

\reference{pkd90} Pinsonneault, M.H., Kawaler, S.D., \& Demarque, P. 1990,
     ApJS, 74, 501

\reference{r78} Rossano, G.S. 1978, AJ, 83, 234

\reference{js93} Schmitt, J.H.M.M., Zinnecker, H., Cruddace, R.,
     \& Harnden, F.R., Jr., 1993, ApJL, 402, L13

\reference{st88} Stahler, S.W. 1988 ApJ, 293, 207

\reference{st97} Stauffer, J. 1997 in {\it Brown Dwarfs and Extrasolar
      Planets}, eds R. Rebolo, E.L. Martin, \& M.R. Zapatero Osorio,
      ASP Conf.\ Series, in press 

\reference{ss94} Strom, K.M. \& Strom, S.E. 1994, ApJ, 424, 237

\reference{sw94} Swenson, F.J., Faulkner, J., Rogers, F.J., \& Iglesias, C.A.
     1994, ApJ, 425, 286

\reference{ve74} Veeder, G.J. 1974, AJ, 79, 1056

\reference{vss} Vrba, F.J., Strom, S.E., \& Strom, K.M. 1976, AJ, 81, 317

\reference{wa86} Walter, F.M. 1986, ApJ, 306, 573
 
\reference{wa92} Walter, F.M. 1992, AJ, 104, 758

\reference{wa88} Walter, F.M., Brown, A., Mathieu, R.D., Myers, P.C.,
     \& Vrba, F.J. 1988, AJ, 96, 297

\reference{wa94} Walter, F.M., Vrba, F.J., Mathieu, R.D., Brown, A., \& Myers,
     P.C. 1994, AJ, 107, 692

\reference{wa87} Walter, F.M., \etal 1987, ApJ, 314, 297

\reference{bw94} Wilking, B., Giblin, T., McCaughrean, M., Rayner, J.,
     Burton, M., \& Zinnecker, H. 1994. in {\it Infrared Astronomy with
     Arrays: The Next Generation}

\reference{bw92} Wilking, B.A., Greene, T.P., Lada, C.J., Meyer, M.R.,
     \& Young, E.T. 1992, ApJ, 397, 520

\end{references}
\end{document}